\documentclass[twocolumn,english]{revtex4-1}
\usepackage{amsmath}
\usepackage{amsthm}
\usepackage{amssymb}
\usepackage{graphicx}
\def\togli#1{}
\def\comment#1{}
\def\labell#1{\label{#1}}
\def\>{\rangle}\def\<{\langle}

\makeatletter
\usepackage{braket}
\usepackage{amsmath}

\makeatother

\usepackage{babel}
\begin{document}

\title{Tight entropic uncertainty relations for systems with dimension
  three to five}

\author{Alberto Riccardi, Chiara Macchiavello and Lorenzo Maccone}

\affiliation{Dip. Fisica and INFN Sez. Pavia, University of Pavia, via Bassi 6,
I-27100 Pavia, Italy}
\begin{abstract}
  We consider two (natural) families of observables $O_k$ for systems
  with dimension $d=3,4,5$: the spin observables $S_x$, $S_y$ and
  $S_z$, and the observables that have mutually unbiased bases as
  eigenstates. We derive tight entropic uncertainty relations for
  these families, in the form $\sum_kH(O_k)\geqslant\alpha_d$, where
  $H(O_k)$ is the Shannon entropy of the measurement outcomes of $O_k$
  and $\alpha_d$ is a constant. We show that most of our bounds are
  stronger than previously known ones.  We also give the form of the
  states that attain these inequalities.
\end{abstract}
\maketitle

Entropic uncertainty relations \cite{key-5,Kraus,key-6} express the
concept of quantum uncertainty nicely since their lower bound is
typically state-independent, in contrast to the Heisenberg-Robertson
ones \cite{key-1,key-2}. The most used one is the Maassen-Uffink
relation \cite{key-6},
\begin{equation}
H(A)+H(B)\geqslant-2\log_{2}c=q_{MU},\label{MU eur}
\end{equation}
where $H(A)$ and $H(B)$ are the Shannon entropies of the measurement
outcomes of two observables $A$ and $B$, and
$c=\max_{j,k}\left|\braket{a_{j}|b_{k}}\right|$ is the maximum overlap
between their eigenstates. It is a state-independent bound, meaningful
even if the observables share some common eigenstates. The bound
\eqref{MU eur} is tight if $A$ and $B$ have mutually unbiased bases
(MUBs) as eigenstates. Stronger bounds for arbitrary observables,
which involve the second largest term in
$\left|\braket{a_{j}|b_{k}}\right|$, have been found recently in
\cite{ColesPiani} and \cite{RPZ}.  If one considers more than two
observables, tight bounds were proven only in few cases, most of them
in dimension $d=2$. For a complete set of MUBs the strongest bounds
were derived by Ivanovic in \cite{Ivanovic} for odd $d$, and by
Sanchez in \cite{Sanchez} for even $d$. Moreover, some bounds for an
incomplete set of MUBs are in \cite{Azarchs}.

In this paper we derive tight entropic uncertainty relations for more
than two observables for systems of dimensions $d=3,4,5$, both for
spin observables and for arbitrary numbers of MUBs. On one hand, for
spin observables we find
\begin{eqnarray}
H(S_x)+H(S_y)+H(S_z)\geqslant\gamma_s
\label{spin3}
\end{eqnarray}
with $\gamma_s=2,3-\tfrac34\log_23\simeq 3.62,3.12$ for spin
$s=1,\tfrac32,2$ respectively. These inequalities are all stronger
than previously known results. The case of $s=1$ has been derived
analytically, while the rest numerically. For half-integer spins the
inequality is saturated by any of the eigenstates of the three spin
observables, while for integer spins the inequality is saturated only
by null projection states. Moreover, we find
\begin{eqnarray}
  H(S_j)+H(S_k)\geqslant
  \xi_s
\label{spin2}\;
\end{eqnarray}
for all $j,k=x,y,z$ ($j\neq k$) with $\xi_s=1,1.71,1.56$ for spin
$s=1,\tfrac32,2$. The case $s=1$ coincides with \eqref{MU eur}, but
the other cases are stronger than previous results. On the other hand,
for observables $\{A_j\}$ with MUBs as eigenstates (the eigenvalues
are irrelevant for EURs) we find, for dimension $d=3$ (where up to four
MUBs exist):
 \begin{eqnarray}
  H(A_1)+  H(A_2)+  H(A_3)&\geqslant& 3,\ \\
  H(A_1)+ H(A_2)+ H(A_3)+H(A_4)&\geqslant& 4;\
\labell{tre}\;
\end{eqnarray}
for dimension $d=4$ (where up to five MUBs exist):
\begin{eqnarray}
  H(A_1)+  H(A_2)+  H(A_3)&\geqslant& 3,\ \\
  H(A_1)+  H(A_2)+  H(A_3)+H(A_4)&\geqslant& 5,\\
  H(A_1)+  H(A_2)+  H(A_3)+H(A_4)+H(A_5)&\geqslant& 7;\ 
\labell{quattro}\;
\end{eqnarray}
and, finally, for dimension $d=5$:
\begin{eqnarray}
  H(A_1)+  H(A_2)+  H(A_3)&\geqslant& 2\log_25\\
  H(A_1)+  H(A_2)+  H(A_3)+H(A_4)&\geqslant& 6.34\\
   {\textstyle{\sum_{j=1}^5}}H(A_j)&\geqslant& 8.33\\
 {\textstyle{\sum_{j=1}^6}}H(A_j)&\geqslant& 10.25\;.
\labell{cinque}\;
\end{eqnarray}

In addition to the above bounds, we also provide the form of the
states that saturate them and we compare them to previous results in
the literature. 

The paper is organized as follows. In Sec.~\ref{s:spin} we consider
spin observables. The case $s=1$ is developed analytically from a
recent parametrization of the state \cite{8}, while the other cases
are solved numerically. In Sec.~\ref{s:mubs} we consider the
observables with MUBs as eigenstates: after a brief review of the
previous results, we derive tight entropic uncertainty relations
through numerical methods. In all cases, we detail the classes of
states that saturate the obtained relations.  In the appendix, we give
the details of the numerical procedures we employed.

\section{Entropic Uncertainty Relations for spin observables}
\labell{s:spin}
We start by considering the entropic uncertainty relations (EUR)
relative to the spin observables $S_x$, $S_y$ and $S_z$ for systems of
different dimensions.
\subsection{Spin $1$}
The state of a three-dimensional system can be written in terms of
$S_x$, $S_y$ and $S_z$ as \cite{8}
\begin{equation}
\rho=\sum_{j=x,y,z}\left(\omega_{j}\left(\mathbb{I}-S_{j}^{2}\right)+\frac{a_{j}S_{j}+q_{j}Q_{j}}{2}\right),\label{Qutrit density matrix}
\end{equation}
where $Q_{j}$ is the anti-commutator of $S_{k}$ and $S_{l}$, with
$j\neq k,l$, i.e.  $Q_{j}=\left\{ S_{k},S_{l}\right\} $, and 
\begin{equation}
\omega_{j}=1-\braket{S_{j}^{2}},\begin{aligned}a_{j}=\braket{S_{j}},\end{aligned}
\begin{aligned}q_{j}=\braket{Q_{j}},\end{aligned}
\label{omega (1)}
\end{equation}
with $0\leq\omega_{j}\leq1$ and $\lvert a_{j}\rvert\leq1.$ In matrix form
\eqref{Qutrit density matrix} is
\begin{equation}
\rho=\left(\begin{array}{ccc}
\omega_{x} & \frac{-ia_{z}-q_{z}}{2} & \frac{ia_{y}-q_{y}}{2}\\
\frac{ia_{z}-q_{z}}{2} & \omega_{y} & \frac{-ia_{x}-q_{x}}{2}\\
\frac{-ia_{y}-q_{y}}{2} & \frac{ia_{x}-q_{x}}{2} & \omega_{z}
\end{array}\right).\label{density matrix}
\end{equation}
The condition Tr$\left[\rho\right]=1$ implies
\begin{equation}
\omega_{x}+\omega_{y}+\omega_{z}=1.\label{Vincolo su gli omega}
\end{equation}
Since $\rho$ is positive-semidefinite, all principal minors of the
right-hand-side of \eqref{density matrix} are non-negative, which
implies the three inequalities $4\omega_{k}\omega_{l}\geqslant
a_{j}^{2}$, for $k,l=x,y,z$ and $j\neq k$, $j\neq l$. These
inequalities can be expressed also as
\begin{equation}
-2\sqrt{\omega_{k}\omega_{j}}\leq a_{j}\leq2\sqrt{\omega_{k}\omega_{l}}.\label{Secondo vincolo sui aj}
\end{equation}

\begin{figure}
{\includegraphics[scale=0.4]{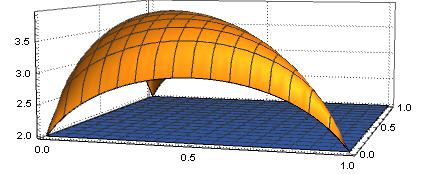}}
\caption{\label{f:fig}
Plot of the function $\Gamma\left(\omega_{x},\omega_{y}\right)$.}
\end{figure}

In the representation where $S_{j}^{2}$ are diagonal, the
spin components are
\begin{eqnarray}
&&S_{x}=\left(\begin{array}{ccc}
0 & 0 & 0\\
0 & 0 & -i\\
0 & i & 0
\end{array}\right),\ 
S_{y}=\left(\begin{array}{ccc}
0 & 0 & i\\
0 & 0 & 0\\
-i & 0 & 0
\end{array}\right),\\&&
S_{z}=\left(\begin{array}{ccc}
0 & -i & 0\\
i & 0 & 0\\
0 & 0 & 0
\end{array}\right).
\end{eqnarray}
The eigenstates of $S_{x}$ are then given by:
\begin{equation}
\begin{aligned}\ket{S_{x}=0}=\left(\begin{array}{c}
1\\
0\\
0
\end{array}\right)\end{aligned}
,\begin{aligned}\ket{S_{x}=\pm1}=\frac{1}{\sqrt{2}}\left(\begin{array}{c}
0\\
\mp i\\
1
\end{array}\right)\end{aligned}
,
\end{equation}
and similar relations for the other observables. The probabilities of
$S_{j}$ are then given by
\begin{equation}
p_{m=0}=\omega_{j},
\ p_{m=\pm1}=\frac{1}{2}\left(1-\omega_{j}\mp a_{j}\right),
\end{equation}
whence one can calculate the Shannon entropies of $S_{j}$ as
\begin{align}
  &H\left(S_{j}\right)= -\frac{1}{2}\left(1-\omega_{j}+a_{j}\right)\log_{2}\left[\frac{1}{2}\left(1-\omega_{j}+a_{j}\right)\right] \\
  &
  -\frac{1}{2}\left(1-\omega_{j}-a_{j}\right)\log_{2}\left[\frac{1}{2}\left(1-\omega_{j}-a_{j}\right)\right]
  -\omega_{j}\log_{2}\omega_{j}.  \nonumber\end{align} For two
observables we find the optimal EUR (\ref{MU eur}),
\begin{equation}
H\left(S_{i}\right)+H\left(S_{j}\right)\geqslant1,\label{MU spin 1}
\end{equation}
indeed $c=\frac{1}{\sqrt{2}}$ and moreover the above inequality is tight when
calculated on the 
null projection state of any of the two observables. For three
observables we obtain an EUR by finding an upper bound to
$-\sum_{j}H\left(S_{j}\right)$. 
To this aim, we can use the conditions (\ref{Secondo vincolo sui aj}),
employing the monotonicity of the logarithm as
\begin{alignat}{1}
&\frac{1}{2}\left(1-\omega_{j}\pm a_{j}\right)\log_{2}\left[\frac{1}{2}\left(1-\omega_{j}\pm a_{j}\right)\right]\\&
\leq\frac{1}{2}\left(1-\omega_{j}+2\sqrt{\omega_{k}\omega_{l}}\right)\log_{2}\left[\frac{1}{2}\left(1-\omega_{j}+2\sqrt{\omega_{k}\omega_{l}}\right)\right].
\nonumber\end{alignat}
Then we have
\begin{align}
&-\sum_{j}H\left(S_{j}\right)\leq \sum_{j}\omega_{j}\log_{2}\omega_{j}\nonumber \\
 &
 +\left(1-\omega_{j}+2\sqrt{\omega_{k}\omega_{l}}
\right)\log_{2}\left[\frac{1-\omega_{j}+2\sqrt{\omega_{k}\omega_{l}}}{2}\right].
\end{align}
The right-hand side  is a function
$-\Gamma\left(\omega_{x},\omega_{y}\right)$ 
which depends only $\omega_x$ and $\omega_y$ since the $\omega_{j}$s
are constrained by (\ref{Vincolo su gli omega}). Inverting the
inequality, we find the EUR
\begin{equation}
{\textstyle{\sum_j}}
H\left(S_{j}\right)\geqslant\Gamma\left(\omega_{x},\omega_{y}\right)\geqslant
2.
\label{eur qutrit spin}
\end{equation}
The lower bound $\Gamma$ is plotted in Fig.~\ref{f:fig}. Its minimum
value $\Gamma=2$ is found for $\omega_{j}=1$ and
$\omega_{k}=\omega_{l}=0.$ These conditions imply that $a_{j}=0$
for all $j$ through (\ref{Secondo vincolo sui aj}).  Thus, $\sum_j
H\left(S_{j}\right)=2$, is attained on null projection states.

This result shows a different behavior of the EUR for spin observables
in the case of integer spin with respect to the half-integer case. A
simple example of the latter is the qubit case: it was shown in
\cite{Sanchez} that for qubits we have $\sum
H\left(S_{j}\right)\geqslant2$, but the minimum is achieved by any of
the eigenstates of one of the $S_{j}$ in contrast to the qutrit case
obtained here. This difference in behavior between integer and
half-integer spins is true also for larger spin numbers (see below).\\

A straightforward generalization of (\ref{MU spin 1}) is obtained by
repeating that inequality for pairs of observables, obtaining $\sum
H\left(S_{j}\right)\geqslant\frac{3}{2}$. It is weaker than our bound
\eqref{eur qutrit spin}.

\subsection{Spin $\frac{3}{2}$}\label{s:spin32}

For a four-dimensional system, we are unaware of a representation of
the density matrix in terms of the spin observables and we cannot
reproduce the derivation given for $s=1$.  We thus develop a simple
computational method that gives tight EUR for small system dimensions
$d$. 

An arbitrary pure state $|\psi\>$ of a $d$-dimensional system depends
on $2d-2$ real parameters. It is sufficient to consider pure states
because of the concavity of the Shannon entropy: mixed states have
greater entropy. The probability on $|\psi\>$ of the measurement
outcomes is $p\left(a_{k}\right)=\left|\braket{a_{k}|\psi}\right|^{2}$
for an arbitrary observable $A=\sum_{k}a_{k}\ket{a_{k}}\bra{a_{k}}$,
whence the entropy is
$H\left(A\right)=\sum_{k}-p\left(a_{k}\right)\log_{2}
\left[p\left(a_{k}\right)\right]$.  Considering $n$ observables
$A_{1},A_{2},..A_{n}$ we can calculate the quantity
$\sum_{j=1}^{n}H\left(A_{j}\right)$, which can be seen as a function
of the $2d-2$ parameters representing the state. This function can
then be numerically minimized over this parameter space.  In addition
to finding the minimum, we then also find the states that saturate the
bounds, which are then tight.  In the Appendix we give more details on
the computational procedure, here we present only the results.

For the case of two spin observables we find
\begin{equation}
H\left(S_{j}\right)+H\left(S_{k}\right)\geqslant1.71,\label{4d 2Spin}
\end{equation}
with $j,k=x,y,z$ and $j\neq k$.

To compare this result with the previous results of \cite{ColesPiani}
and \cite{RPZ}, we can express these as \cite{Review}
\begin{eqnarray}
&&H(A)+H(B)\geqslant\max(q_{CP},q_{RPZ}),\mbox{ with }\\
&&q_{CP}=2\left[-\log_{2}c+\frac{1}{2}\left(1-\sqrt{c}\right)\log_{2}\frac{c}{c_{2}}\right],
\\
&&q_{RPZ}=2\left[-\log_{2}c-\log_{2}\left(b^{2}+\frac{c_{2}}{c}\left(1-b^{2}\right)\right)\right],
\label{RPZ}
\end{eqnarray}
where $b=\frac{1+\sqrt{c}}{2}$,
$c=\max_{j,k}\left|\braket{a_{j}|b_{k}}\right|$ is the maximum overlap
among eigenstates of $A$ and $B$, and $c_2$ is the second maximum
overlap. Both $q_{CP}$ and $q_{RPZ}$ are greater than $q_{MU}$ of
\eqref{MU eur}. Our result (\ref{4d 2Spin}) is an even stronger bound
than both $q_{CP}$ and $q_{RPZ}$. Indeed for $s=1$ we have
$c=\frac{1}{2}\sqrt{\frac{3}{2}}$ and $c_{2}=\frac{1}{2\sqrt{2}}$, so
$q_{CP}=1.59$ and $q_{RPZ}=1.68$.

The bound \eqref{4d 2Spin} is not saturated by one of the eigenstates of
$S_{j}$, indeed for any eigenstate we have
$H\left(S_{j}\right)+H\left(S_{k}\right)\geqslant1.81.$ Instead, it is
saturated by the state
\begin{equation}
\ket{\psi}=\sin\left(15^\circ\right)\ket{0}+\cos\left(15^\circ\right)\ket{2},
\end{equation}
and by similar superpositions weighted by the angle
$\alpha=15^\circ.$ The bound (\ref{4d 2Spin}) is in agreement with
the numerical bound found in \cite{RPZ}, but here we find also the 
state that achieves the minimum.

For the case of three spin observables we find
\begin{equation}
H\left(S_{x}\right)+H\left(S_{y}\right)+H\left(S_{z}\right)\geqslant3-\frac{3}{4}\log_{2}3=3.62.\label{4D Eur spin}
\end{equation}
If we employ (\ref{4d 2Spin}) to obtain a bound for three observables,
(by applying it to each pair of observables) we find
\begin{equation}
H\left(S_{x}\right)+H\left(S_{y}\right)+H\left(S_{z}\right)\geqslant\frac{3}{2}\cdot(1.71)=2.56,
\end{equation}
which is weaker than (\ref{4D Eur spin}). The same argument applied to
$q_{RPZ}$ of \eqref{RPZ} leads to
$H\left(S_{x}\right)+H\left(S_{y}\right)+H\left(S_{z}\right)\geqslant\frac{3}{2}\cdot1.68=2.52$:
also in this case our result \eqref{4D Eur spin} is stronger than
previous ones.

The lower bound \eqref{4D Eur spin} is achieved by the eigenstates of
any of three observables $S_{j}$. This generalizes the result found by
Sanchez in \cite{Sanchez}: indeed in this case the MUBs represent also
the spin components. As mentioned above, the EUR for half-integer and
integer spin values are attained for different classes of states.

\subsection{Spin $2$}
A spin $2$ system has dimension $d=5$. Using the same algorithm
detailed in the previous section, we find
\begin{align}
&H\left(S_{j}\right)+H\left(S_{k}\right)\geqslant1.56\label{Spin 2 Eur 2 obs}
\\
&H\left(S_{x}\right)+H\left(S_{y}\right)+H\left(S_{y}\right)\geqslant3.12.\label{Spin 2 EUR 3 obs}
\end{align}
Both the above inequalities are saturated by the eigenstates
corresponding to the eigenvalue $0$ of any of the three observables
$S_j$, the null projection state (as in the case $s=1$). For example,
 the above inequalities are saturated by the state 
\begin{equation}
\ket{S_{x}=0}=\frac{1}{2}\sqrt{\frac{3}{2}}\ket{0}-
\frac{1}{2}\ket2+\frac{1}{2}\sqrt{\frac{3}{2}}\ket4.
\end{equation}

The comparison of (\ref{Spin 2 Eur 2 obs}) with the previously known
bounds $q_{MU}$, $q_{CP}$ and $q_{RCZ}$ shows that, again, our result
is stronger. In fact, in this case we have
$c=\frac{1}{2}\sqrt{\frac{3}{2}}$ and $c_{2}=\frac{1}{2}$.  Therefore,
$q_{MU}=1.41$, $q_{CP}=1.48$ and $q_{RPZ}=1.53$, which are weaker than
(\ref{Spin 2 Eur 2 obs}). Instead, the numerical bound found in
\cite{RPZ} agrees with ours, but we also provide the states that
saturate it. If we consider the application of (\ref{Spin 2 EUR 3
  obs}) to three spin observables we would obtain
$H\left(S_{x}\right)+H\left(S_{y}\right)+H\left(S_{y}\right)\geqslant\frac{3}{2}\cdot1.56=2.34,$
which is weaker than (\ref{Spin 2 EUR 3 obs}): the three-observable
bound is again stronger than the ones obtained by joining
two-observable bounds.

\section{Entropic uncertainty relations for arbitrary numbers of MUBs}
\label{s:mubs}
We now consider the EURs relative to observables that have mutually
unbiased bases (MUBs) as eigenstates (the eigenvalues are irrelevant
for the EURs). To obtain the EURs we use the same procedure detailed
in Sec.~\ref{s:spin32}. However, we must also calculate the MUBs for
each dimension. In a $d$-dimensional Hilbert space there exist $d+1$
MUBs if $d$ is a power of a prime, otherwise only three bases are
known to exist \cite{DURT}.  The proprieties of MUBs strongly depend
on the dimension, e.g.~for a qubit, MUBs are also the eigenbases of
the spin observables, but this is not true for $d>2$.  The problem of
finding MUBs can be translated into finding Hadamard matrices: the
columns of such matrices are the states of the MUBs. This
problem was solved in \cite{MUBs} for dimensions $d=2,3,4,5$.  Here we
use that result to study EURs: for each dimension $d=3,4,5$ we
consider up to $d+1$ observables $A_{1},A_{2},...,A_{d+1}$ that have
MUBs as eigenstates.

We now briefly review previous results for EURs with MUBs observables.
For any number $L$ of these observables, we can construct an EUR with
a trivial generalization of Maassen and Uffink's relation (\ref{MU
  eur}) by applying \eqref{MU eur} to pairs of bases, obtaining
\begin{equation}
  \sum_{i=1}^LH(A_{i})\geqslant\frac{L}{2}\log_{2}d.\label{Generalization of MU EUR}
\end{equation}
However, this inequality is almost never tight. A better bound was
given in \cite{Ivanovic} for $L=d+1$:
\begin{equation}
{\textstyle{\sum_{i=1}^L}}H(A_{i})\geqslant\left(d+1\right)\left(\log_{2}\left(d+1\right)-1\right)=q_{I},\label{Ivanovic}
\end{equation}
which is also not always tight. For even dimension $d$, a stronger
bound was given in \cite{Sanchez}:
\begin{equation}
\sum_{i=1}^LH(A_{i})\geqslant\left(\frac{d}{2}\log_{2}\frac{d}{2}+\frac{d+1}{2}\log_{2}\frac{d+1}{2}\right)=q_{S},\label{Sanchez}
\end{equation}
which is tight only in dimension two. For $L<d+1$ in
\cite{BallesterWehner} it has been shown that if the Hilbert space
dimension is a square, that is $d=r^{2},$ then for $L<r+1$ the
inequality (\ref{Generalization of MU EUR}) is tight, namely
\begin{equation}
\sum_{i=1}^LH(A_{i})\geqslant\frac{L}{2}\log_{2}d=q_{BW}.\label{Ballester}
\end{equation} A further bound for $L<d+1$
was given in \cite{Azarchs}:
\begin{equation}
\sum_{i=1}^LH(A_{i})\geqslant-L\log_{2}\left(\frac{d+L-1}{d\cdot L}\right)=q_{A}.\label{Archas}
\end{equation}
For more details on the above bounds, we refer to \cite{Review}.  We
now present our results which are tight for all dimensions and all
numbers $L$ of MUBs.

\subsection{Dimension Three}

In dimension $d=3$ four MUBs exist $A_{1}$, $A_{2}$, $A_{3}$ and $A_{4}$, whose states  are respectively given by the columns of
the Hadamard matrices
\begin{align}
&\begin{array}{cc}
M_{1}=\left(\begin{array}{ccc}
1 & 0 & 0\\
0 & 1 & 0\\
0 & 0 & 1
\end{array}\right), &M_{2}= \frac{1}{\sqrt{3}}\left(\begin{array}{ccc}
1 & 1 & 1\\
1 & \omega & \omega^{2}\\
1 & \omega^{2} & \omega
\end{array}\right),\end{array}\\
&\begin{array}{cc}
M_{3}=\frac{1}{\sqrt{3}}\left(\begin{array}{ccc}
1 & 1 & 1\\
\omega^{2} & \omega & 1\\
1 & \omega & \omega^{2}
\end{array}\right), & M_{4}=\frac{1}{\sqrt{3}}\left(\begin{array}{ccc}
1 & 1 & 1\\
\omega & \omega^{2} & 1\\
1 & \omega^{2} & \omega
\end{array}\right),\end{array}\nonumber
\end{align}
with $\omega=\exp\left(\frac{2\pi i}{3}\right)$.  If the system is
prepared in an eigenstate of any of the MUBs, the entropy of that
observable is null while the other entropies are maximal: e.g.~if
$H(A_{1})=0$, then we have $H(A_{1})+H(A_{2})+H(A_{3})=2\log_{2}3$. In
contrast to the qubit case $d=2$, this is {\em not} the state that
gives the strongest EUR for $d=3$. Indeed, the state
$\frac{1}{\sqrt{2}}\left(\ket1-\ket2\right)$ has entropies for all
MUBs equal to $1$: $H(A_{i})=1$.  Therefore,
\begin{align}
&H(A_{1})+H(A_{2})+H(A_{3})\geqslant3\label{3d 3Mubs}\\
&H(A_{1})+H(A_{2})+H(A_{3})+H(A_{4})\geqslant4.\label{3d 4mubs}
\end{align}
We have numerically shown that the above inequalities are the optimal
ones. In addition to the above state, they are saturated also by the
following states
\begin{align}
&\frac{e^{i\varphi}\ket0+\ket1}{\sqrt{2}},\ 
\frac{e^{i\varphi}\ket0+\ket2}{\sqrt{2}},\ 
\frac{e^{i\varphi}\ket1+\ket2}{\sqrt{2}},
\end{align}
where $\varphi=\frac{\pi}{3},\pi,\frac{5\pi}{3}$. Our bound \eqref{3d
  3Mubs} is stronger than (\ref{Generalization of MU EUR}), which in
this case gives
$H(A_{1})+H(A_{2})+H(A_{3})=\frac{3}{2}\log_{2}3=2.38$. For $L=3$ the
bound (\ref{Archas}) gives $q_{A}=2.54$, that is also
weaker than (\ref{3d 3Mubs}).  For a complete set of MUBs $L=4$, the
bound (\ref{Ivanovic}) gives $q_{I}=4$ and is then equal to our
relation \eqref{3d 4mubs}.  However, here we have proven that (\ref{3d
  4mubs}) is a tight relation for $d=3$, and we have provided the
states achieve the minimum.

\subsection{Dimension Four}

In dimension $d=4$ five MUBs exist, whose states are given by the columns of
the Hadamard matrices {\footnotesize{}
\begin{align}
&M_{1}=\left(\begin{array}{cccc}
1 & 0 & 0 & 0\\
0 & 1 & 0 & 0\\
0 & 0 & 1 & 0\\
0 & 0 & 0 & 1
\end{array}\right),\\&
\begin{array}{cc}
M_{2}=\frac{1}{2}\left(\begin{array}{cccc}
1 & 1 & 1 & 1\\
1 & 1 & -1 & -1\\
1 & -1 & -1 & 1\\
1 & -1 & 1 & -1
\end{array}\right), & M_{3}=\frac{1}{2}\left(\begin{array}{cccc}
1 & 1 & 1 & 1\\
1 & 1 & -1 & -1\\
-i & i & i & -i\\
i & -i & i & -i
\end{array}\right),\end{array}
\nonumber\\&\nonumber
\begin{array}{cc}
M_{4}=\frac{1}{2}\left(\begin{array}{cccc}
1 & 1 & 1 & 1\\
i & -i & i & -i\\
-1 & -1 & 1 & 1\\
i & -i & -i & i
\end{array}\right), & M_{5}=\frac{1}{2}\left(\begin{array}{cccc}
1 & 1 & 1 & 1\\
i & -i & i & -i\\
i & i & -i & i\\
-1 & -1 & 1 & 1
\end{array}\right).\end{array}
\end{align}
}Since $d=4=r^2$ is a square for $r=2$, then for $L<r+1=3$ the
inequality \eqref{Generalization of MU EUR} is tight
\cite{BallesterWehner}. This is then the best bound up to $L=2$.
However, for $d=4$ we can consider  up to $L=5$. For example, for
$L=3$ we find that the optimal bound is
\begin{equation}
H(A_{1})+H(A_{2})+H(A_{3})\geqslant3.\label{4D 3MUbs}
\end{equation}
It is achieved by the four states
\begin{equation}
\left(\ket0\pm\ket1\right)\sqrt{2},\ 
\left(\ket2\pm\ket3\right)/\sqrt{2}.
\end{equation}
By symmetry, similar relations holds by permuting the MUBs
observables, but involving the superposition of different eigenstates.
For example $H(A_{1})+H(A_{2})+H(A_{4})\geqslant3$ has lower bound
achieved by $\left(\ket{0}\pm\ket{2}\right)/\sqrt{2}$
and $\left(\ket1\pm\ket3\right)/\sqrt{2}$.

In the case of $L=4$ observables we find
\begin{equation}
H(A_{1})+H(A_{2})+H(A_{3})+H(A_{4})\geqslant5\label{4d 4MUbs}
\end{equation}
as the optimal bound, which is saturated by the states
\begin{alignat}{1}
(\ket0\pm\ket1)/{\sqrt{2}},\; 
{(\ket0\pm\ket2)}/{\sqrt{2}},\;
{(\ket0\pm i\ket3)}/{\sqrt{2}},\\
({\ket1\pm i\ket2})/{\sqrt{2}},\nonumber \;
({\ket1\pm\ket3})/{\sqrt{2}},\;
({\ket2\pm\ket3})/{\sqrt{2}}.\nonumber 
\end{alignat}
Compare our bound (\ref{4d 4MUbs}) to (\ref{Ballester}) and
\eqref{Archas}: for $L=4$ we find $q_{BW}=4$ and $q_{A}=4.77$.
Therefore, our inequality is stronger than both. 

In the case of $L=5=d+1$ observables (the complete set of MUBs), we find
\begin{equation}
H(A_{1})+H(A_{2})+H(A_{3})+H(A_{4})+H(A_{5})\geqslant7,\label{4D 5MUBs}
\end{equation}
which is saturated by states of the following form:
\begin{equation}
\ket{\psi_{jk}}=\frac{1}{\sqrt{2}}\left(\ket j\pm\left(i\right)^{t}\ket k\right),
\end{equation}
with $t=0,1$ and $j$ and $k$ are the eigenstates of $A_{1}$.  For
$d=4$ the inequality (\ref{Sanchez}) gives
$q_{S}=2+\frac{5}{2}\log_{2}5=5.30$, so that (\ref{Sanchez}) is weaker
than our bound (\ref{4D 5MUBs}) in this case.

\subsection{Dimension Five}
In dimension $d=5$ six MUBs exist, whose states are given by the columns of
the Hadamard matrices {\footnotesize{}
\begin{alignat}{1}
 M_{1}=\left(\begin{array}{ccccc}
1 & 0 & 0 & 0 & 0\\
0 & 1 & 0 & 0 & 0\\
0 & 0 & 1 & 0 & 0\\
0 & 0 & 0 & 1 & 0\\
0 & 0 & 0 & 0 & 1
\end{array}\right),\\
M_{2}=\frac{1}{\sqrt{5}}\left(\begin{array}{ccccc}
1 & 1 & 1 & 1 & 1\\
1 & \omega & \omega^{2} & \omega^{3} & \omega^{4}\\
1 & \omega^{2} & \omega^{4} & \omega & \omega^{3}\\
1 & \omega^{3} & \omega & \omega^{4} & \omega^{2}\\
1 & \omega^{4} & \omega^{3} & \omega^{2} & \omega
\end{array}\right),\\
M_{3}=\frac{1}{\sqrt{5}}\left(\begin{array}{ccccc}
1 & 1 & 1 & 1 & 1\\
\omega & \omega^{2} & \omega^{3} & \omega^{4} & 1\\
\omega^{4} & \omega & \omega^{3} & 1 & \omega^{2}\\
\omega^{4} & \omega^{2} & 1 & \omega^{3} & \omega\\
\omega & 1 & \omega^{4} & \omega^{3} & \omega^{2}
\end{array}\right),\\
M_{4}=\frac{1}{\sqrt{5}}\left(\begin{array}{ccccc}
1 & 1 & 1 & 1 & 1\\
\omega^{3} & \omega^{4} & 1 & \omega & \omega^{2}\\
\omega^{2} & \omega^{4} & \omega & \omega^{3} & 1\\
\omega^{2} & 1 & \omega^{3} & \omega & \omega^{4}\\
\omega^{3} & \omega^{2} & \omega^{3} & 1 & \omega^{4}
\end{array}\right),
\\
M_{5}=\frac{1}{\sqrt{5}}\left(\begin{array}{ccccc}
1 & 1 & 1 & 1 & 1\\
\omega^{2} & \omega^{3} & \omega^{4} & 1 & \omega\\
\omega^{3} & 1 & \omega^{2} & \omega^{4} & \omega\\
\omega^{3} & \omega & \omega^{2} & \omega^{2} & 1\\
\omega^{2} & \omega & 1 & \omega^{4} & \omega^{3}
\end{array}\right),\\
M_{6}=\frac{1}{\sqrt{5}}\left(\begin{array}{ccccc}
1 & 1 & 1 & 1 & 1\\
\omega^{4} & 1 & \omega & \omega^{2} & \omega^{3}\\
\omega & \omega^{3} & 1 & \omega^{2} & \omega^{4}\\
\omega & \omega^{4} & \omega^{2} & 1 & \omega^{3}\\
\omega^{4} & \omega^{3} & \omega^{2} & \omega & 1
\end{array}\right),
\end{alignat}}
with $\omega=\exp\left(\frac{2\pi i}{5}\right)$.
For three MUBs observables we find that the optimal bound is
\begin{equation}
H(A_{1})+H(A_{2})+H(A_{3})\geqslant2\log_{2}5,\label{5d 3mubs}
\end{equation}
which is saturated by any eigenstate of any of the three MUBs, as in
the qubit $d=2$ case (also there the EUR for three complementary
observables is saturated by the eigenstates of the observables).  The
bound \eqref{5d 3mubs} is the only known entropic uncertainty relation, apart from the qubit case, 
with more than two observables that has this property. In this
respect, it is somewhat similar to Maassen and Uffink's \eqref{MU
  eur}: they are both achieved by the eigenstates of one of the
observables (so that the entropies of the others are maximum).  For
$L=3$ in (\ref{Archas}) we have $q_{A}=3.30$ while
$2\log_{2}5=4.64.$ Our bound is stronger than these also in this case.

For four MUBs we find that the optimal bound is
\begin{equation}
H(A_{1})+H(A_{2})+H(A_{3})+H(A_{4})\geqslant6.34,\label{5d 4mubs}
\end{equation}
and the minimum is achieved by states that are superposition of four
basis states, e.g.
\begin{equation}
\ket{\psi}=0.19e^{i\frac{5}{3}\pi}\ket0+0.19\ket1+0.68e^{i\frac{9}{5}\pi}\ket3+0.68\ket4.\label{stato 5d}
\end{equation}
In this case we have $q_{A}=5.28$, that is again weaker than our bound \eqref{5d
  4mubs}.  For five MUBs we find
\begin{equation}
H(A_{1})+H(A_{2})+H(A_{3})+H(A_{4})+H(A_{5})\geqslant8.33\;.\label{5d 5mubs}
\end{equation}
and, finally, for the complete set of six MUBs we find
\begin{equation}
\sum_{i=1}^{6}H(A_{i})\geqslant 10.25\;.\label{5d 6mubs}
\end{equation}
The two above inequalities are again minimized by states that can be
expressed by the superposition of four basis states, having the same
form of (\ref{stato 5d}). For $L=5$ we can compare \eqref{5d 5mubs} to
\eqref{Archas} which gives a weaker bound $q_{A}=7.34$, while for the
complete set of MUBs we can compare \eqref{5d 6mubs} to
(\ref{Ivanovic}), which gives a weaker bound $q_{I}=9.51$.

\section{Conclusions}

In this paper we have found several tight entropic uncertainty
relations for two classes of observables: the spin observables $S_x$,
$S_y$, $S_z$ and the observables $\{A_j\}$ with MUBs eigenstates.

For the case of spin observables, for $s=1$ we found a tight relation
(\ref{eur qutrit spin}) for the complete set of spin observables, its
minimum value is achieved by null projection states of any of three
observables. The same types of states saturate also the inequality
(\ref{Spin 2 EUR 3 obs}) for the case of $s=2$. Instead, in the case
$s=\frac{3}{2}$ the inequality (\ref{4D Eur spin}) is minimized by
eigenstates of any of three spin observables.  For both
$s=\frac{3}{2}$ and $s=2$ we have also found tight inequalities for
two spin observables, which are in agreement with the optimal bound
found in \cite{RPZ}, and we have given the states that minimize them.
In the case of $s=2$
they are the null projections states. 

For the case of MUBs observables, we have derived several tight
inequalities for dimensions $d=3,4,5$.  For $d=3$ the results (\ref{3d
  4mubs}) equals the previous bound (\ref{Ivanovic}) but here we also
found the class of states that saturates it. In contrast, for $d=4,5$,
the bounds (\ref{4D 5MUBs}) and (\ref{5d 6mubs}) represent stronger
EUR than known ones. New inequalities have been also found for
incomplete sets of MUBs in every dimension: in each case the new
bounds are tight and we have derived the states that achieve the
minimum. We note the peculiar behavior of (\ref{5d 3mubs}), which is
achieved by any eigenstate of one of the three MUBs, resembling the
behavior of qubit systems.

\appendix

\section{Numerical methods}
Here we detail the numerical methods used to derive most of our
entropic uncertainty relations. We have used the software package
Mathematica.  For the sake of illustration, we consider the case of
$s=\frac{3}{2}.$ The most general pure state of a quantum system for
$d=4$ is
\begin{alignat}{1}
\ket{\psi}= & e^{i\chi_{0}}\sin a_{0}\sin a_{1}\cos a_{2}\ket0+e^{i\chi_{1}}\sin a_{0}\sin a_{1}\sin a_{2}\ket1+\nonumber \\
 & +e^{i\chi_{2}}\sin a_{0}\cos a_{1}\ket2+\cos a_{0}\ket1
\end{alignat}
where $a_{i}\in\left[0,\frac{\pi}{2}\right]$ and
$\chi_{0}\in\left[0,2\pi\right].$ To compute the probability
distributions of $S_{x},S_{y}$ and $S_{z}$ over the state $\ket{\psi}$
we work in the representation of eigenstates of $S_z$. In this
representation, the spin matrices are 
\begin{align}
&S_{x}=\frac{1}{2}\left(\begin{array}{cccc}
0 & \sqrt{3} & 0 & 0\\
\sqrt{3} & 0 & 2 & 0\\
0 & 2 & 0 & \sqrt{3}\\
0 & 0 & \sqrt{3} & 0
\end{array}\right),
\\&
S_{y}=\frac{1}{2i}\left(\begin{array}{cccc}
0 & \sqrt{3} & 0 & 0\\
-\sqrt{3} & 0 & 2 & 0\\
0 & -2 & 0 & \sqrt{3}\\
0 & 0 & -\sqrt{3} & 0
\end{array}\right),
\\&
S_{z}=\frac{1}{2}\left(\begin{array}{cccc}
3 & 0 & 0 & 0\\
0 & 1 & 0 & 0\\
0 & 0 & -1 & 0\\
0 & 0 & 0 & -3
\end{array}\right).
\end{align}
The probability distribution of $S_{z}$ is 
\begin{align}
&p\left(S_{z}=+2\right)=\sin^{2}a_{0}\sin^{2}a_{1}\cos^{2}a_{2};
\\&
p\left(S_{z}=+1\right)=\sin^{2}a_{0}\sin^{2}a_{1}\sin^{2}a_{2};
\\&
p\left(S_{z}=-1\right)=\sin^{2}a_{0}\cos^{2}a_{1};
\\&
p\left(S_{z}=-2\right))=\cos^{2}a_{0}.
\end{align}
Then the entropy is $H\left(S_{z}\right)=
-\sum_{l}p_{l}\left(S_{z}=l\right)\log_{2}p_{l}\left(S_{z}=l\right),$
which depends only on the three parameters $a_j$.\\
To calculate the entropy for $S_{x}$, consider its eigenstates
\begin{align}
&\ket{S_{x}=\pm2}=\frac{1}{2\sqrt{2}}\left(\ket0\pm\sqrt{3}\ket1-\sqrt{3}\ket2\pm\ket3\right);
\\&
\ket{S_{x}=\pm1}=\frac{1}{2\sqrt{2}}\left(\sqrt{3}\ket0\pm\ket1-\ket2\mp\sqrt{3}\ket3\right).\nonumber
\end{align}
We can compute the probability distribution of $S_{x}$ over
$\ket{\psi}$ with
\begin{equation}
p\left(S_{x}=\pm l\right)=\left|\braket{\psi|S_{x}=\pm l}\right|^{2},
\label{probabilty Sx}
\end{equation}
This expression depends on all six parameters of $\ket{\psi}$, and we
can use it to calculate the entropy
$H\left(S_{x}\right)=-\sum_{l}p_{l}\left(S_{x}=l\right)\log_{2}p_{l}\left(S_{x}=l\right).$
\\
An analogous procedure can be used for $S_{y}$, whose eigenstates are
\begin{align}
&\ket{S_{y}=\pm2}=\frac{1}{2\sqrt{2}}\left(\ket0\pm i\sqrt{3}\ket1-\sqrt{3}\ket2\mp i\ket3\right);\\&
\ket{S_{y}=\pm1}=\frac{1}{2\sqrt{2}}\left(\sqrt{3}\ket0\pm i\ket1+\ket2\pm i\sqrt{3}\ket3\right),
\end{align}
whence we can calculate the probabilities and the entropy.

To obtain the optimal EUR, we need to minimize the sum of two or three
of above entropies.  Due to their non linear dependence on the
parameters, it is highly nontrivial to find the minimum analytically.
We have therefore resorted to numerical methods: Mathematica permits
the minimization of a function $f(x_{1},...,x_{n})$ that depends on
$n$ parameters with the routine
\begin{equation}
NMinimize\left[\left\{ f(x_{1},...,x_{n}),\gamma(x_{1},..,x_{n})\right\} ,\left\{ x_{1},..,x_{n}\right\} \right],
\end{equation}
where $\gamma$ represents possible constraints. This routine returns
both the minimum value of the function and also the parameter values
that attain it, which in our case identify the states that minimize
the EUR. For example, if we define
\begin{equation}
f\left(a_{0},a_{1},a_{2},\chi_{0},\chi_{1},\chi_{2}\right)=H\left(S_{x}\right)+H\left(S_{z}\right),
\end{equation}
the instruction
\begin{equation}
NMinimize\left[f\left(a_{i},\chi_{i}\right),\left\{ a_{0},a_{1},a_{2},\chi_{0},\chi_{1},\chi_{2}\right\} \right],
\end{equation}returns
\begin{equation}
\left\{ 1.71,\left\{ a_{0}\rightarrow\frac{\pi}{12},a_{1}\rightarrow\frac{\pi}{4},a_{2}\rightarrow\text{\ensuremath{\frac{\pi}{4}}},\chi_{1}\rightarrow\pi\right\} \right\} ,
\end{equation}
which implies (\ref{Spin 2 Eur 2 obs}). The other relations we
derived can be similarly obtained.

For example, for the case of spin $2$ we can repeat the above
procedure. Again, we can choose the representation of eigenbasis of
$S_{z}$ which  gives
\begin{align}
S_{x} & =\frac{1}{2}\left(\begin{array}{ccccc}
0 & 2 & 0 & 0 & 0\\
2 & 0 & \sqrt{6} & 0 & 0\\
0 & \sqrt{6} & 0 & \sqrt{6} & 0\\
0 & 0 & \sqrt{6} & 0 & 2\\
0 & 0 & 0 & 2 & 0
\end{array}\right),\\
S_{y}&=\frac{1}{2i}\left(\begin{array}{ccccc}
0 & 2 & 0 & 0 & 0\\
-2 & 0 & \sqrt{6} & 0 & 0\\
0 & -\sqrt{6} & 0 & \sqrt{6} & 0\\
0 & 0 & -\sqrt{6} & 0 & 2\\
0 & 0 & 0 & -2 & 0
\end{array}\right),
\\
S_{z} & =\left(\begin{array}{ccccc}
2 & 0 & 0 & 0 & 0\\
0 & 1 & 0 & 0 & 0\\
0 & 0 & 0 & 0 & 0\\
0 & 0 & 0 & -1 & 0\\
0 & 0 & 0 & 0 & 1
\end{array}\right).
\end{align}

\end{document}